\documentclass[a4paper,10pt]{amsart}
\usepackage{graphicx}
\usepackage{amssymb,amsmath,amstext,amscd}

\theoremstyle{plain}
\newtheorem{thm}{Theorem}[section]

\newtheorem*{main theorem}{Theorem}

\theoremstyle{definition}

\begin{document}

\title{Relativity on two-dimensional spacetimes}
\author{Do-Hyung Kim}
\address{Department of Mathematics, College of Natural Science, Dankook University,
San 29, Anseo-dong, Dongnam-gu, Cheonan-si, Chungnam, 330-714,
Republic of Korea} \email{mathph@dankook.ac.kr}


\begin{abstract}
Lorentz transformation on two-dimensional spacetime is obtained
without assumption of linearity. To obtain this, we use the
invariance of wave equations, which is recently proved to be
equivalent to the causality preservation.

\end{abstract}

\maketitle

\section{Introduction} \label{section:1}

Einstein's special relativity begins with two postulates. The
first is the principle of relativity, which states that all
physical laws are the same for any inertial observers. The second
postulate is the constancy of the speed of the light. From these
two postulates, Einstein obtained Lorentz transformation as the
following form.

\begin{eqnarray*}
x^\prime &=& \frac{x-vt}{\sqrt{1-\frac{v^2}{c^2}}} \\ y^\prime &=&
y
\\ z^\prime &=& z \\ t^\prime &=&
\frac{t-\frac{v}{c^2}x}{\sqrt{1-\frac{v^2}{c^2}}}
\end{eqnarray*}

However, in Einstein's original paper, he assumed that our
universe is homogeneous and isotropic to guarantee that the
desired spacetime coordinate transformation is linear, and then he
obtained the desired coordinate transformation.

In contrast to this, in Ref. \cite{Wave},  it is shown that, when
$n \geq 3$, the principle of the constancy of the speed of the
light or equivalently, invariance of wave equations implies the
spacetime coordinate transformation $\verb"x"^\prime = a A
\verb"x" +\verb"b"$ where $A$ is a Lorentz matrix and $a$ is a
positive real number. Then, by use of the principle of relativity
, we can determine $a$ to be 1. In conclusion, we can obtain
Lorentz transformation from the two postulates of relativity
without any assumption on linearity, homogeneity and isotropy.

However, it turns out that this does not hold in the case $n = 2$,
since when $n=2$, the principle of the constancy of the speed of
light gives us many more non-linear candidates for spacetime
coordinate transformations. Compared to the case $n \geq 3$, the
constancy of the speed of the light does not imply the linearity
of the spacetime coordinate transformation in the case $n=2$.

Therefore, it is a natural question to ask what the roles of the
first postulate and the second postulate are. In this paper, this
differences are discussed and eventually, we obtain the desired
coordinate transformations by the two postulates, though their
roles are different in the case $n \geq 3$ and $n=2$.

\section{Review on High-dimensional spacetimes} \label{section:2}

In this section, we review the mathematically rigorous derivation
of Lorentz transformation.

For convenience, we set the speed of the light to be 1, and then,
mathematically, the principle of the constancy of the speed of the
light can be described as the following.

$$\sum\limits_{i=1}^{n-1} \frac{\partial^2 \varphi}{\partial x_i^2} -
\frac{\partial^2 \varphi}{\partial t^2} = 0 \,\,\, \mbox{if and
only of } \,\,\, \sum\limits_{i=1}^{n-1} \frac{\partial^2
\varphi}{\partial x_i^{\prime2}} - \frac{\partial^2
\varphi}{\partial t^{\prime2}} = 0,$$

 where $(x_i, t)$ is the coordinate system of observer $S$ and
 $(x_i^\prime, t^\prime)$ is the coordinate system of observer
 $S^\prime$.

 In \cite{Wave}, it is shown that the above condition is equivalent to the
 equation
  $$(x_1^\prime, \cdots, x_{n-1}^\prime, t^\prime)^t = aA
(x_1, \cdots, x_{n-1}, t)^t + (b_1, \cdots, b_{n+1}),$$ where $a$
is a positive real number and $A$ is a Lorentz matrix that
preserves time-orientation. This is a general form of causal
automorphisms on Minkowski spacetime $\mathbb{R}^n_1$ with $n \geq
3$.(Ref. \cite{Zeeman}). In other words, Einstein's second
 postulate means preservations of causal relations on spacetimes.
 To determine the positive constant $a$, we apply Einstein's first
 postulate. If an observer $S^\prime$ moves with velocity $v$
 relative to $S$, then the velocity of $S$ relative to $S^\prime$
 must be $-v$. By use of this, we can show that the constant $a$
 must be 1 and so, after all, we can obtain the Lorentz
 transformation from the invariance of wave equations(i.e. the principle of
 the constancy of the speed of the light) and the
 principle of relativity.

\section{Two-dimensional spacetimes} \label{section:3}

 We now study spacetime coordinate transformations on
 two-dimensional spacetimes and let us assume that an inertial
 observer $S$ uses his coordinate system $(x,t)$ and observer
 $S^\prime$ uses his coordinate system $(X,T)$.

 Recently, the following Theorem has been proved.(Ref.
 \cite{Kim}, \cite{CQG3} and \cite{CQG4}).

 \begin{thm}
For any $C^2$ function $f$, $\frac{\partial^2 f}{\partial x^2} -
\frac{\partial^2 f}{\partial t^2} = 0 \Leftrightarrow
\frac{\partial^2 f}{\partial X^2} - \frac{\partial^2 f}{\partial
T^2} = 0$ with $\frac{\partial T}{\partial t} > 0$, if and only if
there are two homeomorphisms $\varphi$ and $\psi$ on the real
line, which are either both increasing or both decreasing such
that, if $\varphi$ and $\psi$ are increasing, then we have $X =
\frac{1}{2} \varphi(x+t) + \frac{1}{2}\psi(x-t)$ and $T =
\frac{1}{2}\varphi(x+t) - \frac{1}{2}\psi(x-t)$, and if $\varphi$
and $\psi$ are both decreasing, then we have $X =
\frac{1}{2}\varphi(x-t)+\psi(x+t)$ and $T = \frac{1}{2}
\varphi(x-t) - \frac{1}{2}\psi(x+t).$

\end{thm}

In the above Theorem, the condition $\frac{\partial T}{\partial t}
> 0$ means that the spacetime coordinate transformation must
preserve the time-orientation, which is natural in physical sense.

Compared with the high-dimensional case, it seems that the
principle of the constancy of the speed of the light gives us many
more candidates for spacetime coordinate transformation on
two-dimensional spacetime. However, if we now apply the principle
of relativity, we can obtain a simple form.

Without loss of generality, we assume that $\varphi$ and $\psi$
are both increasing. We also assume that $\varphi(0)=0$ and
$\psi(0)=0$.

Our goal is to find $\varphi$ and $\psi$ that satisfy

\begin{eqnarray}
X &=& \frac{1}{2} \varphi(x+t) + \frac{1}{2}\psi(x-t) \\
T &=& \frac{1}{2} \varphi(x+t) - \frac{1}{2}\psi(x-t)
\end{eqnarray}

by use of the principle of relativity.

If we assume that the relative velocity of $S^\prime$ with respect
to $S$ is v, then the principle of relativity implies that the
relative velocity of $S$ with respect to $S^\prime$ must be $-v$.

This can be expressed as $\frac{dx}{dt}|_{X=0} = v$ and
$\frac{dX}{dT}|_{x=0} = -v$. From equations (1) and (2), we have

\begin{eqnarray}
v &=& \frac{\psi^\prime(\psi^{-1}(-T)) -
\varphi^\prime(\varphi^{-1}(T))}{\psi^\prime(\psi^{-1}(-T)) +
\varphi^\prime(\varphi^{-1}(T))} \\
-v &=& \frac{\varphi^\prime(t) -
\psi^\prime(-t)}{\varphi^\prime(t)+\psi^\prime(-t)}
\end{eqnarray}

If we rearrange (3) and (4), integration gives us the following.

\begin{eqnarray}
\varphi^{-1}(T) &=& \frac{v+1}{v-1} \psi^{-1}(-T) \\
\varphi(t) &=& \frac{v-1}{v+1} \psi(-t)
\end{eqnarray}

In (5), if we let $T = \varphi(t)$, we obtain $\varphi(t) = -
\psi(\frac{v-1}{v+1}t)$. By equating this with (6), we have
$-\psi(\frac{v-1}{v+1}t) = \frac{v-1}{v+1}\psi(-t)$. If we let $s
= \frac{v-1}{v+1}t$, then we obtain $\psi(s) = \frac{1-v}{1+v}
\psi(\frac{1+v}{1-v}s)$. Likewise, we can show that $\varphi(t) =
\frac{1-v}{1+v} \varphi(\frac{1+v}{1-v}t)$.

In other words, both $\varphi$ and $\psi$ must satisfy the
following functional equation.

$$ f(at) = af(t) \,\,\, \mbox{with} \,\,\, a = \frac{1+v}{1-v}. $$

Since $a >1$, we have $f(0)=0$. By differentiating, we have
$f^\prime(at)=f^\prime(t)$ for all real numbers $t$. If we let
$s=at$, then $f^\prime(s)=f^\prime(\frac{s}{a})$ and thus we have
$f^\prime(s)=f^\prime( \frac{1}{a^n}s)$ for any natural numbers
$n$. Since $f^\prime$ is continuous at $0$, we have
$$ f^\prime(0) = \lim_{n \rightarrow \infty}
f^\prime\big(\frac{1}{a^n}s\big) = f^\prime(s) \,\,\,\, \mbox{for
all} \,\, s.$$

 In other words, $f^\prime$ is constant and so, since $f(0)=0$,
$f$ is a linear function. Therefore, there are two real numbers
$\alpha$ and $\beta$ such that $\varphi(t) = \alpha t$ and
$\psi(t) = \beta t$. Since $\varphi$ and $\psi$ are increasing
functions, both $\alpha$ and $\beta$ must be positive.

In conclusion, we have the following spacetime coordinate
transformations.

\begin{eqnarray*}
X &=& \frac{1}{2}(\alpha + \beta)x + \frac{1}{2}(\alpha - \beta)t
\\
T &=& \frac{1}{2}(\alpha - \beta)x + \frac{1}{2}(\alpha + \beta)t
\end{eqnarray*}

If we consider $\frac{dX}{dT}|_{x=0} = -v$, we have $\beta =
\frac{1+v}{1-v} \alpha$, and finally we have the desired spacetime
coordinate transformations.

\begin{eqnarray}
X &=& \frac{\alpha}{1-v}(x-vt) \\
T &=& \frac{\alpha}{1-v}(t-vx)
\end{eqnarray}

To determine the constant $\alpha$, consider a rod with rest
length $L$. If the road is at rest in the reference frame $S$,
then the transformation (7) and (8) tell us that $S^\prime$
measures the length as $X_2-X_1 = \alpha (1+v)L$. If the rod is at
rest in the reference frame $S^\prime$, then $S$ measures the
length as $x_2-x_1 = \frac{(1-v)L}{\alpha}$. The principle of
relativity ensures that these two lengths must be the same and
then we get $\alpha = \sqrt{\frac{1-v}{1+v}}$.

In conclusion, we have the transformation :

\begin{eqnarray*}
X &=& \frac{x-vt}{\sqrt{1-v^2}} \\
T &=& \frac{t-vx}{\sqrt{1-v^2}}
\end{eqnarray*}

\section{Acknowledgement}

The present research was conducted by the research fund of Dankook
University in 2013.

\end{document}